# Ab-initio investigations of novel potential all-d metal Heusler alloys Co₂MnNb

Sumit Kumar[1,2[0000-0001-7276-8287]], Diwaker[3[0000-0002-4155-7417]], Vivek Kumar[2[0000-0001-7971-0897]], Karan S. Vinayak[1[0000-0002-1380-8370]] and Shyam Lal Gupta[4[0000-0003-0472-3436]]

¹ Department of Physics, DAV College Sec-10, Chandigarh – 160010, India
² Department of Physics, Government College Una, H.P. – 174303, India
³ Department of Physics, SCVB Govt. College, Palampur, Kangra, H. P. – 176061, India
⁴ HarishChandra Research Institute (HRI), Prayagraj, U. P. – 211019, India
shyamlalgupta@hri.res.in

**Abstract.** In this study, we employ the Wien2k code to conduct ab-initio study of a novel potential all-d-metal Heusler alloy Co₂MnNb. The analysis utilizes the comparison of local spin density approximations (LDA) with Perdew-Burke-Ernzerh parameterized Generalized Gradient Approximation (PBE-GGA) for structural optimization while modified Becke-Jones potential (mBJ) exchange-correlation potentials to examine various characteristic properties of the alloy under study. Employing Birch-Murnaghan equation of state, we construct the energy-versus-volume curve, facilitating the determination of stable phases and structural parameters of the investigated alloys. Structural optimization in both non-magnetic (NM) and spin-polarized (FM) states reveals the stability of the alloy in the FM state. The compound exhibits metallic behavior in bulk, with notable anisotropic semiconducting behavior for down spin while pure metallic behavior for up spin electrons. Partial density of states of each element of the composition is also analysed to compare their respective contribution towards the observed band structure. The anisotropic behavior of Co₂MnNb for a specific spin state could be of importance in future spintronic and other thin films device applications.

## 1 Introduction

Ever-since the inception in 1903 Heusler alloys [1] constitute a remarkable materials class encompassing over > 1000 distinct members, each exhibiting a diverse array of extraordinary properties. Their versatility renders them invaluable across various applications, including spintronics [2, 3, 4, 5], shape memory [6, 7], giant magnetoresistance (GMR) [8, 9], and thermo-electric applications [10, 11, 12, 13]. Despite nearly a century of intensive study, Heusler alloys remain a subject of enduring fascination among scholars, owing to their immense combinatorial potential and wide range of extraordinary properties including half metallicity; ferro and ferri magnets at room temperature and at high temperature. Half metals act as metals for one(up/down) spin type while as insulators or semiconductors for opposite(down/up) spin orientation.

Some materials in half metals exhibit anisotropic behavior for a specific spin. In different directions, they behave as a metal, an insulator, or a semiconductor. The search for Heusler alloys and ferromagnetic semiconductors with stable half-metallicity and high curie temperature is still a daunting task for material science researchers. Since half metallic compounds has special properties based upon direction of spin, so it becomes important to understand the relation between structure and properties to design new materials for the desired functionality. To fulfill this need, computational simulations plays a pivotal role to explore the new compounds and their properties. This will make it easy for experimentalists to narrow down their choices among the wide range of possible materials. The studies of magnetic Heusler compounds designed specifically for spintronic applications paved the way for new possibilities for future. In recent 7-8 years, more than 750 papers are published on full Heusler alloys. Recently, many novel compositions for potential Heusler alloys, $Co_2CuAl$[14], $Co_2NbGe$[15], $Ir_2VZ$ (Z=Sn,In)[16], $Ir_2CrZ$ (Z=Sn,In)[17] were investigated via first principle calculations. Junaid Jami et.al.[18] investigated iron based full heusler alloy $Fe_2MnSn$ in all possible phases. Ambrose et.al.[19] made a groundbreaking investigation on Heusler alloy film growth a few years ago by growing a film of $Co_2MnGe$ on the GaAs(001) substrate. They observed lattice constant enhancement for thin film samples compared to its bulk form. As a result researchers explored $Co_2Mn$ based Heusler alloys for half-matallicity at room temperature. Consequently, Cobalt and Manganese based Heusler alloys becomes center of attraction for researchers to realize room and high temperature half- metallicity. Paula and Reis [20, 21] explained in detail that replacing the p-block element (Z in $X_2YZ$ Heusler alloy) with a transition element increases mechanical ductility and induces large reversible mechano-caloric effects. They highlighted that replacing p-d hybridization with d-d hybridization improves the alloy's mechanical characteristics. Some compositions shows magnetism with high $T_c$ values and following the Slater-Pauling rule. As a result, this effect may broaden the range of spintronic applications. They are commonly known as all-d-metal [21] Heusler alloys. The improved functional behavior reveals them constituting new class of materials for promising state-of-the-art technological applications. In this study, the properties of an all-d-metal novel Heusler alloy $Co_2MnNb$ by first principles using WIEN2K[22] DFT code in fully ordered $L2_1$ structure in possible antiferromagnetic (AFM) and ferromagnetic configuration(FM).

## 2 Computational Details

The electronic and magnetic behavior along with the structural properties of $Co_2MnNb$ are reported in this work. The method of full-potential linearized augmented plane-wave (FP-LAPW)[23, 24, 25] implemented within a commercially available DFT [26, 27] based software package named as WIEN2k is used in the present study. We have employed the PBE-GGA and LDA to model and thence to compare these electronic exchange–correlations [28, 29] on structural properties. Fur-ther to analyze the the band structures(electronic) and electronic DOS more accurately, we have utilized the modified Becke Jones (mBJ) exchange and correlation potential[30]. A k-point grid of 10x10x10 k-mesh distributed over the Brillouin zone (BZ), value of 9.0 for product $RMT \times K_{max}$, and that of 2.18, 2.18, and 2.13 (a.u) are used as non-overlapping muffin-tin radii for Co, Mn, and Nb, respectively. While $G_{max}$ is set as 12, the energy cutoff (core) of -8.0 Ryd was chosen to define the separation of valence and core states. The convergence limit of 10-4 Ry is used for performing these self consistent calculations. The ground state energy for the ferromagnetic (FM) phase and all possible antiferromagnetic structures are investigated in L2$_1$ structure. Furthermore, we also report the fraction of spin polarization (SP) in percentage as given by[31, 32]

$$\textbf{SP} = \frac{\eta \uparrow - \eta \downarrow}{\eta \uparrow + \eta \downarrow} \times \textbf{100} \qquad (1)$$

where $\eta\uparrow$ and $\eta\downarrow$ represent the majority and minority spin states.

## 3    Results and Discussion

### 3.1    Structural Properties

To determine the ground state energy of Co$_2$MnNb, we have evaluated non-magnetic (NM), antiferro-magnetic (AFM) alongwith ferro-magnetic (FM) phases in cubic L2$_1$ structure. For the AFM phase, we investigated all six possible spin configurations for ground state energy: udu (Co-up, Mn-down, and Nb-up), dud, uud, ddu, udd, and duu. The conventional unit cells of the above phases employed to carry out the computations is depicted in Fig. 1. X-crysden package [33] is used to visualize the crystal structure.

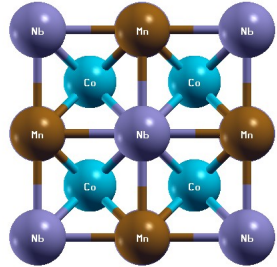

**Fig. 1**. Co$_2$MnNb in L2$_1$ structure (Fm-3m space group)

PBE-GGA and LDA exchange-correlation potentials are used to investigate the ground state energy and volume optimization in NM, AFM and FM phases. Birch Murunaghan equation of state is then used to fitting the variations of obtained total energy of the system with its volume and thence to get the optimized values of vari-ous parameters including the lattice constant as given in Table – 1. Total Energy vs Volume curve of $Co_2MnNb$ alloy in PBE-GGA and LDA for NM and FM phases are shown in Fig. 2 and Fig. 3 respectively. Among all the investigated phases, FM phase in $L2_1$ structure is the energetically most favorable with lattice constant of 5.9723A° with -15532.2116 Ryd. This structure with the lowest energy spin configuration is then further investigated for other properties.

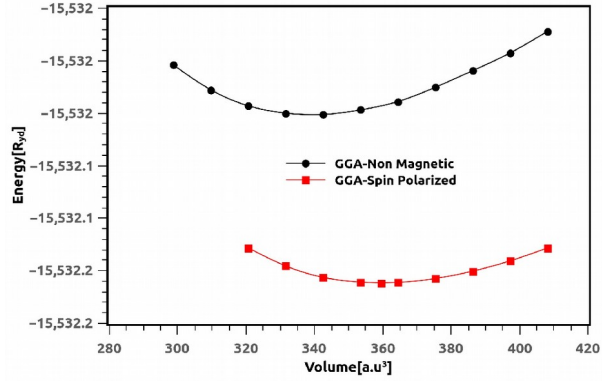

**Fig. 2.** Volume optimization in GGA

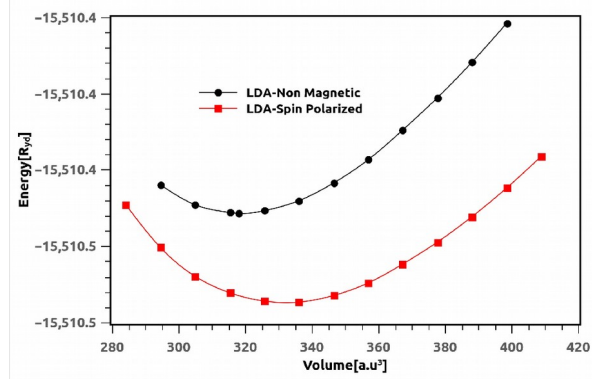

Fig. 3. Volume optimization in LDA

The formation energy of $Co_2MnNb$ alloy is further predicted as following for analyzing its thermodynamic stability,

$$\Delta H_f = E_{tot} - (2xE_{Co} + E_{Mn} + E_{Nb}) \qquad (2)$$

where the term $E_{tot}$ represents the total energy of ground state of the compound $Co_2MnNb$ per formula unit and the terms $E_{Co}$, $E_{Mn}$ and $E_{Nb}$ represent the total ground state energies for the bulk Co, Mn and Nb respectively. Using the calculated values of respected energies in equation-2, we found $\Delta H$ as $-1.51 ev/f.u.$ which indicates its phase stability.

**Table 1.** Calculated parameters for $Co_2MnNb$ in different phases

| Phase (XC Potential) | Structure | Bulk modulus (GPa) | Latt. Constant (A°) | Tot. Energy (Ry) | Volume (a.u³) |
|---|---|---|---|---|---|
| FM (GGA) | L2₁ | 188.4249 | 5.9723 | -15532.2116 | 359.3856 |
| NM (GGA) | L2₁ | 232.7926 | 5.8547 | -15532.0474 | 338.5679 |
| FM (LDA) | L2₁ | 220.1766 | 5.8105 | -15510.5369 | 330.9571 |
| NM (LDA) | L2₁ | 272.7349 | 5.7344 | -15510.4782 | 318.1289 |
| AFM-udu (GGA) | L2₁ | 185.3116 | 5.9754 | -15532.2114 | 359.9545 |
| AFM-dud (GGA) | L2₁ | 183.3883 | 5.9727 | -15532.2111 | 359.4508 |
| AFM-uud (GGA) | L2₁ | 185.9791 | 5.9745 | -15532.2115 | 359.7883 |
| AFM-ddu (GGA) | L2₁ | 186.3019 | 5.9734 | -15532.2115 | 359.5896 |
| AFM-udd (GGA) | L2₁ | 186.0180 | 5.9737 | -15532.2115 | 359.6339 |
| AFM-duu (GGA) | L2₁ | 187.8176 | 5.9730 | -15532.2115 | 359.5162 |

## 3.2 Electronic and Magnetic Properties

Fig. 4 and Fig. 5 depict the calculated total and projected DOS for the FM phase within the L2₁ structure while Fig. 6 and Fig. 7 show the respective band structures(electronic) for majority as well as minority spins. The asymmetric distribution among those spins depicts the magnetic nature and exhibit lack of complete spin polarization at at the Fermi energy $E_F$. The projected DOS (pDOS) plots (see Fig. 4 and Fig. 5) establish Co and Mn as primary contributors compare to Nb having negligible contribution towards total DOS. The broad spectrum of d states of majority spins is attributed to hybridization of d-states among Co, Mn and Nb atoms. An analysis of the band structure for the majority spin channel (see Fig. 6) echoes the metallic behavior generated by the fusion of conduction and valence bands. The analysis of minority spins (see Fig. 7) indicates a semiconducting gap at zone center Γ and along symmetry direction L, while a slight merger of valence and conduction bands around the rest of the symmetry directions, leading to a bulk metallic behavior.

As anticipated, the total magnetic moment (see Table-2) in the FM L2₁ structure ($6.0656\mu_B$) predominantly originates from the two symmetrically equivalent Co atoms (each Co atom contributing $1.5391\mu_B$) and the Mn atom (contributing $3.4946\mu_B$), with the Nb atom exhibiting a negligible induced negative magnetic moment. Closer look into Table - 2 indicates that Slater Pauling rule is not being followed in $Co_2MnNb$. The Slater Pauling rule demands $M = (N_V - 24) \mu_B$, where M represents the net magnetic moment while $N_V$ is the total number of valence electrons in a unit cell. Each unit cell in $Co_2MnNb$ has 28 valence electrons.

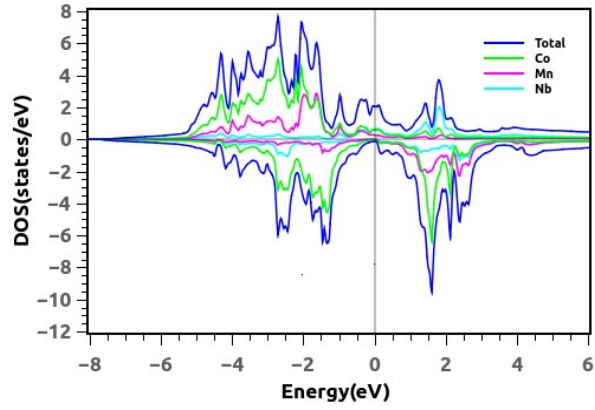

**Fig. 4.** Spin polarized total and projected DOS

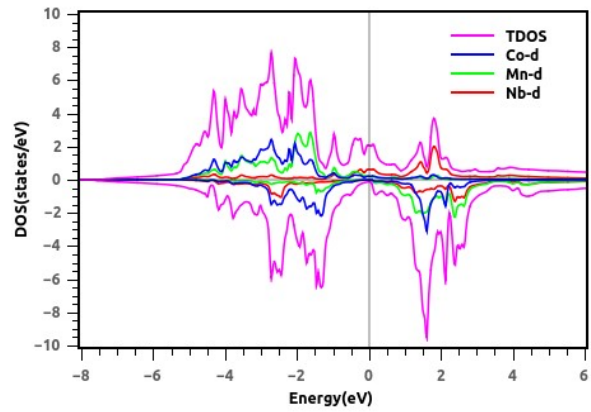

**Fig. 5.** Spin polarized total and d-orbital DOS

The magnetic moment should be 4 (Slater-Pauling rule); however the obtained larger magnetic moment indicates the presence of other interactions namely spin-orbit interactions controlling the magnetic behavior.

## 4 Conclusion

Within the sprawling compositional spectrum of full Heusler alloys, first-principles computations are the most feasible method to predict experimentally realizing specific composition with desired properties. The adopted pathway via DFT based WIEN2K software package in this study entails a step wise identification of the theoretically most stable crystal and magnetic structure. The FM state of the cubic L2$_1$ structure is found to be the energetically most stable with net magnetic moment of 6.056$\mu_B$ per formula unit, primarily originating from Co and Mn constituent atoms. The detailed DOS analysis indicate role of the hybridization of d-orbitals towards the magnetic behavior. The TDOS confirmed the metallic behavior of the compound with spin polarization of 85.67 percent at $E_F$. The DOSs and band structures show that the bulk behaviour of the novel potential Heusler alloy under investigation is metallic for the majority spins and anisotropic behaviour(semiconducting gap at $\Gamma$ center and L direction whereas metallic in other high symmetry directions) for the minority spin channel. This anisotropic $Co_2MnNb$ behavior for a specific spin may be crucial in spintronic devices. Additional investigations into phonon behavior, magnetic anisotropy and exchange interactions along with thermodynamic properties will be required inorder to explore and establish the studied composition as technologically important and viable Heusler alloy.

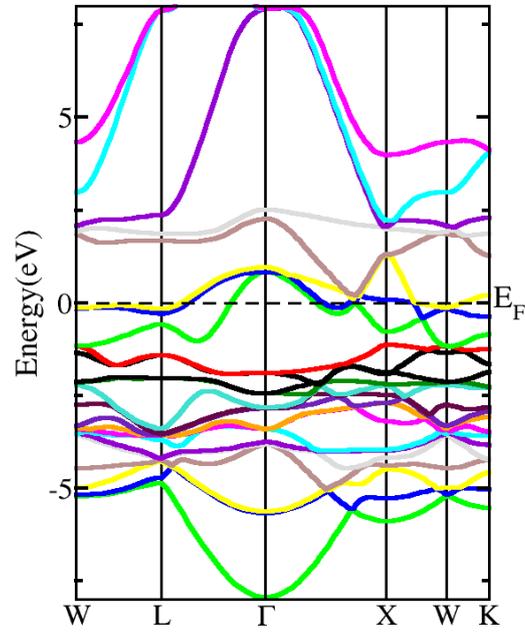

**Fig. 6.** Band structure of Co₂MnNb (Spin-Up)

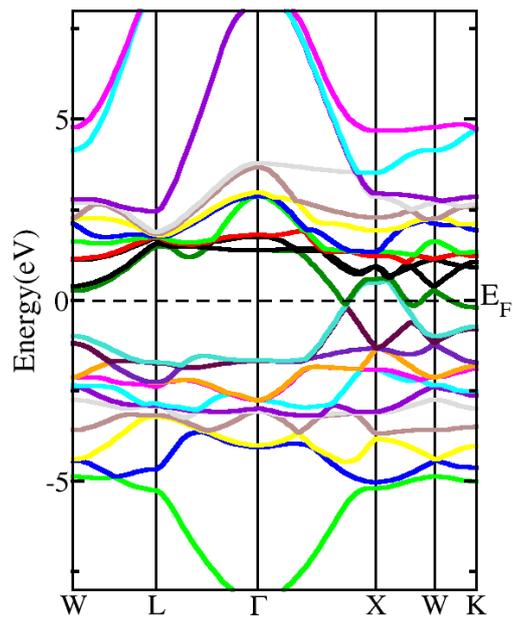

**Fig. 7**. Band structure of Co₂MnNb (Spin-Down)

**Table 2.** Total and partial magnetic moments of $Co_2MnNb$

| Quantity per unit cell | Value in $\mu_B$ |
|:---:|:---:|
| $\mu_{tot}$ | 6.0656 (mBJ) |
| | 5.9503 (GGA) |
| | 5.7358 (LDA) |
| $\mu_{Co}$ | 1.5391 (mBJ) |
| | 1.3245 (GGA) |
| | 1.2472 (LDA) |
| $\mu_{Mn}$ | 3.4946 (mBJ) |
| | 3.3365 (GGA) |
| | 3.0983 (LDA) |
| $\mu_{Nb}$ | -0.1697 (mBJ) |
| | -0.0558 (GGA) |
| | -0.0048 (LDA) |